\documentclass[%
preprint,
superscriptaddress,
 aps,
floatfix,
longbibliography,
]{revtex4-1}

\usepackage{graphicx}
\usepackage{amssymb}
\usepackage{xcolor}
\usepackage[fleqn]{amsmath}
\usepackage{mathtools}
\usepackage{float}
\usepackage{mdframed}
\usepackage{url}
\usepackage{booktabs}
\usepackage{natbib}
\usepackage[colorlinks=true,linkcolor=blue,citecolor= blue]{hyperref}%
\usepackage{xspace}
\usepackage[capitalise]{cleveref}
\usepackage{physics}
\newcommand\hl[1]{%
  \bgroup
  \hskip0pt\color{green!40!black}%
  #1%
  \egroup
}
\newcommand\hlo[1]{%
  \bgroup
  \hskip0pt\color{black!80!black}%
  #1%
  \egroup
}
\newcommand\deleted[1]{%
  \bgroup
  \hskip0pt\color{red!40!black}%
  #1%
  \egroup
}

\begin{document}

\newcommand{\Iv}{$I_{v}$\xspace}
\newcommand{\mumacro}{$\mu$\xspace}
\newcommand{\mumicro}{$\mu_m$\xspace}
\newcommand{\phim}{$\phi_m$\xspace}
\newcommand{\yd}{$\dot\gamma$\xspace}
\newcommand{\muofiv}{$\mu(I_v)$\xspace}
\newcommand{\shear}{$\sigma_\mathrm{shear}$\xspace}

\preprint{APS/123-QED}

% \title{Shear thickening and the role of interparticle friction in the macroscopic friction coefficient of dense suspensions.}
% \title{Shear thickening and the role of interparticle friction in the macroscopic friction coefficient of dense suspensions.}
\title{Shear Thickening of Dense Suspensions: The Role of Friction}

\author{Vishnu Sivadasan}
\affiliation{Computational Science Lab, Institute for Informatics, University of Amsterdam.}
\author{Eric Lorenz}
\affiliation{Computational Science Lab, Institute for Informatics, University of Amsterdam.}
\affiliation{Electric Ant Lab B.V., Amsterdam, The Netherlands}
\author{Alfons G. Hoekstra}
\affiliation{Computational Science Lab, Institute for Informatics, University of Amsterdam.}
\affiliation{ITMO University, Saint Petersburg, Russian Federation.}
\author{Daniel Bonn}
\affiliation{Institute of Physics, Faculty of Science, University of Amsterdam}
% \collaboration{CLEO Collaboration}%\noaffiliation

\date{\today}% It is always \today, today,
             %  but any date may be explicitly specified% Force line 

\begin{abstract}
Shear thickening of particle suspensions is caused by a transition between lubricated and frictional contacts between the particles.
Using 3D numerical simulations, we study how the inter-particle friction coefficient (\mumicro) influences the effective macroscopic friction coefficient (\mumacro)  and hence the microstructure and rheology of dense shear thickening suspensions. We propose expressions for \mumacro in terms of distance to jamming for varying shear stresses and \mumicro values. We find \mumacro to be rather insensitive to interparticle friction, which is perhaps surprising but agrees with recent theory and experiments. 

\end{abstract}

\pacs{Valid PACS appear here}% PACS, the Physics and Astronomy
\maketitle

\section{Introduction}
% intro on shear thickening suspensions
Understanding the rheological properties of shear thickening suspensions is scientifically challenging and highly relevant from  the viewpoint of several applications \cite{lee2003ballistic, li2015shear, decker2007stab, cwalina2016engineering}. The phenomenon of shear thickening~\cite{denn2018shear,Fall2012a,Fall2015,Brown2013,denn2014rheology,cwalina2014material,boyer2011unifying} in which the viscosity increases with increasing shear rate and shear stress, is attributed to the formation of frictional contacts between the particles as suggested by computational results \cite{Mari2014b,Seto2013b, Ness2015e} and confirmed by experiments \cite{Comtet2017,Lin2015,pan2015s,Royer2016,huang2005flow}. 
Shear thickening suspensions can be characterized by their macroscopic friction coefficient \mumacro, given by $\mu = \sigma_\mathrm{shear}/P$, with \shear the shear stress and $P$ the confining pressure. Using suspensions under constant confining pressure, Boyer et al. \cite{boyer2011unifying} demonstrated that \mumacro is  a unique function of a viscous parameter \Iv defined as $I_v=\eta_f \dot\gamma/P$, where $\eta_f$ and $\dot\gamma$ are the fluid viscosity and the shear rate respectively. They observe similar \muofiv behavior for different materials (polystyrene, PMMA) and particle sizes. Gallier et al.\cite{gallier2014rheology} studied \muofiv rheology in simulations for $\phi<0.45$ ($\phi$ being the particle volume fraction) and their simulations agree quantitatively with the experimental results. However, a more detailed analysis of  \mumacro and associated changes in the microstructure of the suspension is needed to shed further light on the behavior of the macroscopic friction coefficient \mumacro and notably its relation with the microscopic inter-particle friction coefficient \mumicro. Here, we  perform 3D numerical simulations of dense shear thickening suspensions with varying inter-particle friction coefficients to study associated changes on \mumacro. 
Based on recent results on constitutive relationships for shear thickening systems \cite{singh2018constitutive,wyart2014discontinuous}, we propose analytic expressions for \mumacro in terms of distance to jamming ($\phi_m-\phi$, where $\phi_m$ is the jamming volume fraction) for constant volume systems with varying pressure, shear stress and \mumicro values. Using the average coordination number as a parameter, the microstructure of the particles in the system is analyzed to assess its influence on \mumacro. Finally, simulations of non-spherical particles are performed to study the effect of non-sphericity on the behavior of the macroscopic friction coefficient.  

% microscopic and macroscopic friction coefficients

% \input{Models_and_Method}
\section{Methods}

The numerical simulations were performed using the simulation framework SuSi \cite{Lorenz2018}. We use  the Lattice Boltzmann Method (LBM) based fluid to simulate the fluid field and Lagrangian particles as the solid phase. The fluid-particle interactions are modelled with the Noble Torczynski Method \cite{nobletorczynski}. Lubrication forces are calculated explicitly at particle gaps smaller than the LBM lattice spacing. Adaptive refinement of timesteps is performed in order to ensure numerical stability and accuracy, as the inter-particle forces diverge at small particle gaps. The contact normal force $\mathbf{F}_\mathrm{rep}$ between particles is calculated from the overlap of a contact repulsion layer \cite{Lorenz2018} of specified thickness $d_c \approx 0.001R$ \cite{Mari2014b}, where $R$ is the mean radius of particles.  
\begin{align}
\mathbf{F}_\mathrm{rep} = \left\{ \begin{array}{ll} - c_0\dfrac{(d-d_\mathrm{c})^2}{d d_{\mathrm{c}}^2}\mathbf{e}_h, & d\le d_\mathrm{c} \\ 
0 & \mathrm{otherwise} \end{array} \right. 
\label{eqn_frep}
\end{align}
where $c_0$ is the repulsion coefficient, $d$ is the gap between the particles, $d_c$ is the repulsion layer thickness and $\mathbf{e}_h$ is the connecting unit vector between the particles.  The static and kinetic friction between particles is modeled as proposed by Luding~\cite{Luding2008a}. Upon initiation of frictional contact between particle pairs, a linear spring of length $\xi$ is initialized between the closest surface points to model static friction and is updated using the relative tangential velocity between the two contacting surface points. The maximum static friction is $F_\mathrm{s} \leq \mu_\mathrm{s} |\mathbf{F}_\mathrm{norm,fric}|$, as given by Coulomb's Law. The spring force $ \mathbf{F}_\mathrm{spr}$ is applied if the amplitude of $\mathbf{F}_\mathrm{spr}=-k\mathbf{\xi}$ is smaller than the maximum possible static friction force $F_\mathrm{s}$. Kinetic friction $F_\mathrm{k} = \mu_\mathrm{k} |\mathbf{F}_\mathrm{norm,fric}|$ is applied as a tangential force at the surface points  if $F_\mathrm{spr}$ exceeds $F_\mathrm{s}$. For kinetic friction, the static friction spring length is rescaled so that $F_\mathrm{spr} = F_\mathrm{k}$. 
% This ensures a smooth transition between static and kinetic friction. 
In our simulations, we keep $\mu_s=\mu_k=\mu_\mathrm{m}$, where \mumicro is referred to as the microscopic friction coefficient.   

The interacting particles are deemed frictional based on a Critical Load Model~\cite{Mari2014b}, where two particles are considered to be in friction if the normal force ($F_\mathrm{rep}$) between the contacting particles exceeds a threshold value ($F_\mathrm{CL}$). The static and kinetic friction is based on the normal force for friction ($F_{\mathrm{norm,fric}}$), calculated as \cite{Mari2014b}:  
\begin{equation}
F_{\mathrm{norm,fric}} = \left\{ \begin{array}{ll} |\mathbf{F}_\mathrm{rep}| - F_\mathrm{CL} & \quad\text{if } |\mathbf{F}_\mathrm{rep}| \geq F_\mathrm{CL}, \\ 0 & \quad\mathrm{otherwise}. \end{array} \right. 
\end{equation}

For the simulations discussed in the subsequent sections, a $96\mu m \times 64 \mu m \times 96 \mu m $ system is used, which contains $\approx 650$ particles for $\phi=0.56$. The particles are have a mean diameter of $8 \mu m$ with a standard deviation of $0.4 \mu m$ to avoid crystallization. 
The particles are neutrally buoyant in the suspending fluid, which mimics water (fluid viscosity $\eta_f=1.002 \times 10^{-3} Pa.s$, density $\rho_f=1000 kg/m^3$). The simulated systems have a characteristic stress for frictional contacts, given by $\sigma_0 = F_\mathrm{CL}/(6\pi R^2)$, where $R$ is the average particle radius. 
\hlo{For the performed analysis, we choose instances of the system with average shear stress greater than $\sigma_0$, so that frictional interactions are significant.}
% The states of the system with stresses larger than $\sigma_0$ are chosen for analysis, in order to ascertain the role of the microscopic friction coefficient.

% \input{Results_and_Discussion}
\section{Results and Discussion}
% \subsection{Comparison with Viscous Number Rheology}
The macroscopic friction coefficient (\mumacro) of suspensions is characterized by the viscous number  (\Iv) of the suspension flow.  \Iv is defined as $I_v = \eta_f \dot\gamma/P$, where $\eta_f$ is the fluid viscosity, \yd is the shear rate and $P$ is the pressure in the system. 
\hlo{The viscous number can be seen as the ratio of the internal timescale of microscopic particle rearrangements in a viscous system ($\eta_f/P$), to the macroscopic flow timescale ($1/\dot\gamma$).}
Boyer et al. \cite{boyer2011unifying}  used pressure imposed flows to study variation in \mumacro with \Iv, where systems of hard spheres were sheared at constant pressure ($P$) and shear rate (\yd) while the system was allowed to dilate (changing $\phi$) in order to keep $P$ constant. They demonstrated that \mumacro of suspensions is the sum of contact ($\mu_c$) and hydrodynamic ($\mu_h$) stress contributions, as shown in Eq.\ref{eq_muofiv}. 
\begin{align}
\mu(I_v) = \underbrace{\mu_1 + \dfrac{\mu_2-\mu_1}{1+I_0/I_v}}_{\mu_c} + \underbrace{I_v + \dfrac{5}{2}\phi_m I_v^{\frac{1}{2}}}_{\mu_h}
\label{eq_muofiv}
\end{align}
\hlo{Where, $\mu_1$ is the limit of of the particle contact contribution to macroscopic friction ($\mu_c$) at vanishing viscous numbers,  
and $\mu_2$ is the maximum $\mu_c$ at $I_v \rightarrow \infty$ as observed in granular flows \cite{cassar2005submarine,jop2015rheological}. $I_0$ represents the scale over which $\mu_c(I_v)$ changes and is observed to be constant for a given particle shape. \phim is the jamming volume fraction.
} 
$\mu_h(I_v)$ is designed to reproduce the Einstein viscosity at low $\phi$ and be non-saturating at high \Iv.
Here, simulations of constant $\phi$ and \yd with varying $P$ are performed to study \muofiv. In this study, we define $P$ as the average of the diagonal elements of the stress tensor in the system i.e. $P = \sum_{i=1}^{3}\sigma_{ii}/3$. We systematically vary the microscopic friction coefficient \mumicro and compare to the predictions of \muofiv rheology (\cref{eq_muofiv}), to see if the constant $\phi$ and \yd simulations conform to the predictions of \muofiv rheology.
\begin{figure}[h]
\centering
\includegraphics[width=160mm]{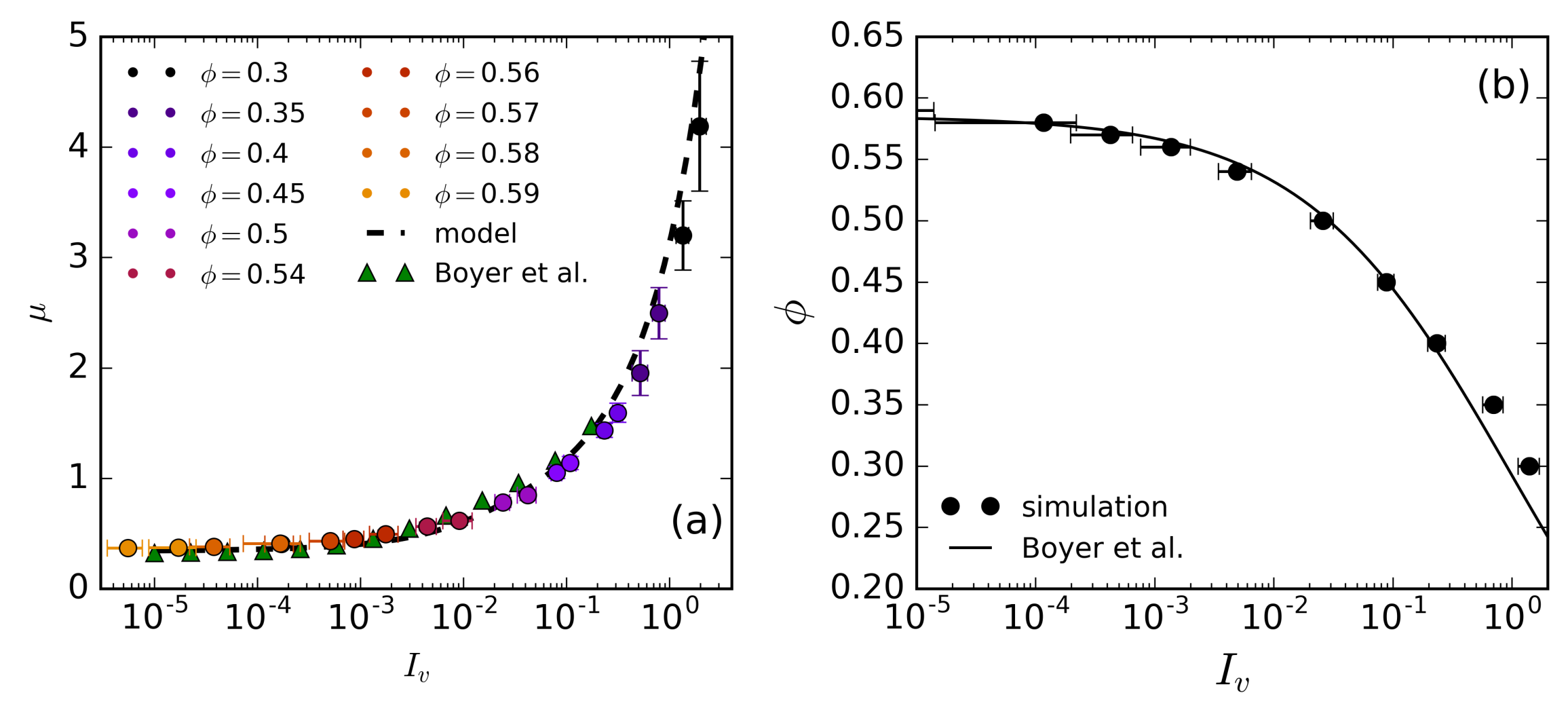}
% \decoRule
\caption{ \textbf{(a)} Macroscopic friction coefficient \mumacro vs viscous number \Iv  comparison between simulation and model. Dots represents \mumacro prediction from simulations of $\phi$ corresponding to its color. 
% The points cover a range of shear rates ($1.0/s \leq \dot\gamma \leq 1000.0/s$), and only points where the shear stress of the system $\sigma_\mathrm{shear}$ is larger than the onset stress $\sigma_0$ is used in the analysis in order to avoid  frictionless states (equivalent to \mumicro=0).
The dashed line shows the \muofiv prediction from \cref{eq_muofiv} with $\mu_1=0.34$ (minimum \mumacro observed), $\mu_2=0.7$ and $I_0=0.009$ providing a good fit to the simulation results. The microscopic friction coefficient $\mu_m=0.5$. Triangles represent the experimental results from Boyer et al.~\cite{boyer2011unifying}. Vertical and horizontal errorbars correspond to variation in \mumacro and \Iv in the data, in each \Iv interval. 
\textbf{(b)} Variation in $\phi$ vs viscous number \Iv. Dots represent simulation results, and the line represents results from Boyer et al. \cite{boyer2011unifying}. Error bars represent the range of \Iv values observed for a given $\phi$.}
\label{fig_mu_vs_Iv_mus1_0}
\end{figure}

Fig.\ref{fig_mu_vs_Iv_mus1_0}(a) compares the the results from our simulations  to the \muofiv rheology predicted by Eq.\ref{eq_muofiv}, and the experimental results from Boyer et al.~\cite{boyer2011unifying}. Suspensions of different $\phi$ values were simulated to obtain the range of \Iv values.  
It can be observed that $\mu \approx 0.34$ at vanishing \Iv, which is similar to the values obtained in experiments \cite{boyer2011unifying,cassar2005submarine}. Using $\mu_2=0.7$ and $I_0\approx0.009$ provides a good fit to the simulation data. The value for $\mu_2$ is the same as that observed previously in experiments and simulations of spherical particles \cite{boyer2011unifying,gallier2014rheology}.

% \hlo{Here, $\mu_2$ represents the contribution from the particle contacts to the macroscopic friction coefficient observed at high  viscous numbers ($I_v>>1$)  \cite{cassar2005submarine,jop2015rheological}, and is close to the values observed in experiments and simulations \cite{boyer2011unifying,gallier2014rheology}}.    

At vanishing \Iv, we find high corresponding $\phi$ values similar to that in experiments \cite{boyer2011unifying}. Under constant $\phi$ settings, the range of \Iv values accessible for each  $\phi$ value is limited (as seen in \cref{fig_mu_vs_Iv_mus1_0}(b)), and multiple simulations of varying $\phi$ values are required to capture \Iv values varying in orders of magnitude. This issue can be overcome by allowing the system to dilate in order to change $\phi$, as done in experiments. The variation in $\phi$ with \Iv is shown in \cref{fig_mu_vs_Iv_mus1_0}(b), along with the experimental observation from Boyer et al.\cite{boyer2011unifying}. The simulations show good agreement with the experimental results.  

\subsection{Effect of varying microscopic friction coefficient}
\begin{figure}[h]
\centering
\includegraphics[width=160mm]{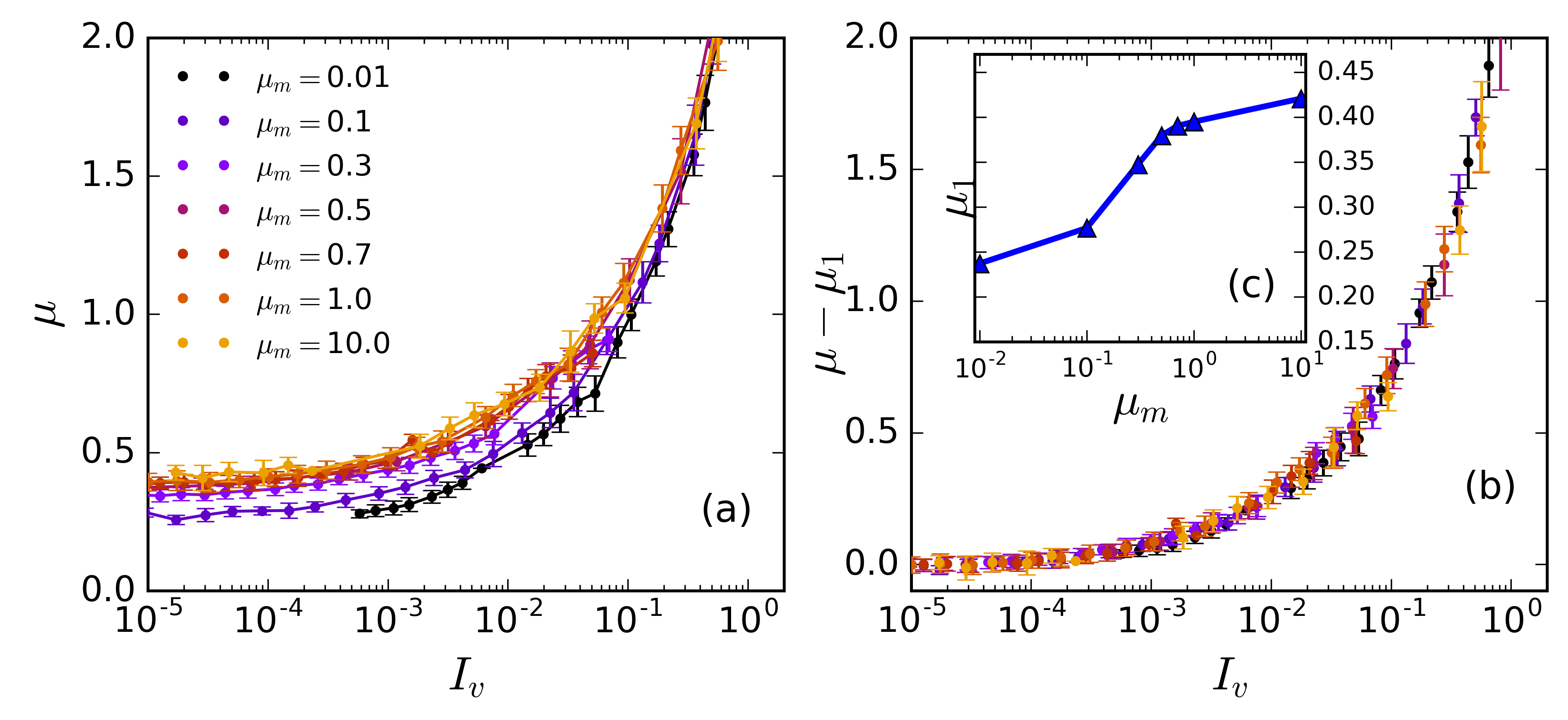}
% \decoRule
\caption{\textbf{(a)} Macroscopic friction coefficient \mumacro vs viscous number \Iv for different  microscopic friction coefficients (\mumicro). Each dot represents the prediction from simulations of corresponding \mumicro value. Results are compiled over various $\phi$ and \yd values for each \mumicro value in consideration.  
\textbf{(b)} $\mu - \mu_1$ vs \Iv where $\mu_1$ is the minimum \mumacro observed. 
\textbf{(c)} Change in the minimum \mumacro observed (i.e. $\mu_1$) with \mumicro.}
\label{fig_mu_vs_Iv_mus_all}
\end{figure}

Earlier simulation studies of the role of the microscopic friction coefficient (\mumicro) were performed at large viscous numbers ($I_v>0.1$) with limited overlap between \Iv ranges studied in experiments \cite{gallier2014rheology}. Here, a larger range of \Iv values is accessed, allowing comparisons with experimental results at lower \Iv values. In order to study the effect of changing \mumicro on \mumacro, simulations of $0.01 \leq \mu_m \leq 10$ are performed, while keeping all other system parameters the same. This amounts to over 500 individual simulations, the results of which are presented in \cref{fig_mu_vs_Iv_mus_all}.  

In \cref{fig_mu_vs_Iv_mus_all}(a) the simulation results of \muofiv for various \mumicro values are shown. At large \Iv values ($I_v>0.1$), \muofiv is similar for all \mumicro values.  At vanishing \Iv values ($I_v<10^{-4}$), the minimum \muofiv (i.e. $\mu_1$) reduces with decreasing \mumicro, as shown in  \cref{fig_mu_vs_Iv_mus_all}(c). This observation is in agreement to that made in past simulations of 2D granular and suspension flows \cite{da2005rheophysics,trulsson2017effect}.  
Interestingly, the relationship between $\mu-\mu_1$ and \Iv  collapses to the same curve for all \mumicro values in this system (see \cref{fig_mu_vs_Iv_mus_all}(b)). Such a collapse was not observed when spherical particle suspensions studied in this section are compared against non-spherical particle suspensions (see \cref{section_nonsperhical_particles}), suggesting that particle shape is a factor here.   
The change in $\mu_1$ with \mumicro follows a sigmoidal relationship, as observed in   \cref{fig_mu_vs_Iv_mus_all}(c).
% The effect of microscopic friction coefficient (\mumicro) on the microstructure is more significant, and is discussed in \cref{section_microstructure}. 
\hlo{
The collapse of $\mu-\mu_1$ for $I_v<10^{-3}$ with the viscous number is obviously due to \mumacro being constant and equal to $\mu_1$ in this range. Within the intermediate viscous number range ($10^{-3}\leq I_v \leq 10^{-1}$) where the particle contact contribution ($\mu_c$ in \cref{eq_muofiv}) to \mumacro is dominant, the variation in \mumacro with the microscopic friction coefficient \mumicro is dictated by the variation in $\mu_2-\mu_1$ with \mumicro. Seeing that $\mu_2$ is rather insensitive to  microscopic inter-particle friction coefficients ($\mu_2$ varies between 0.7 and 0.8 for completely frictionless and frictional particles respectively \cite{gallier2014rheology}), we estimate that the largest difference in $\mu-\mu_1$ between systems of $\mu_m=0.01$ and $\mu_m=10.0$ should be $\approx 0.2$, which agrees with the observed variations in $\mu-\mu_1$ with \mumicro at $I_v \approx 10^{-1}$. For large viscous number range ($I_v>10^{-1}$), the variations in \mumacro are dominated by the hydrodynamic component ($\mu_h$ in \cref{eq_muofiv}), and does not depend on the friction. 

% Although the data for $\mu-\mu_1$ has an apparent collapse for all \mumicro values, some minor variations are observed at $10^{-2} \leq I_v \leq 10^{-1}$ with changing \mumicro. 
% The standard deviations in \mumacro observed in these ranges are of the same magnitude as the expected variations mentioned above, and prevents us from resolving the results sufficiently to detect the subtle differences in $\mu - \mu_1$ with \mumicro. 

% Essentially, $\mu-\mu_1$ encodes the distance of a suspension from jamming and \Iv is the ratio of timescale of micro-scale particle rearrangements in a viscous medium ($\tau_\mathrm{micro}=\eta_f/P$) to the macroscopic shear timescale imposed ($\tau_\mathrm{macro}=1/\dot\gamma$)~\cite{boyer2011unifying}. 
% The main contribution to $\mu-\mu_1$ is therefore from distance to jamming.
The main contribution to $\mu-\mu_1$ is therefore given by the distance to jamming.
The collapse of the data for $\mu-\mu_1$ as a function of \Iv  for $0.01 \leq \mu_m  \leq 10$ implies that at the same microscopic to macroscopic particle rearrangement timescale ratios (i.e. \Iv), all systems will have the same distance to jamming, regardless of their microscopic friction coefficient. 
This also entails that if $\mu-\mu_1$ indeed is a measure of the distance of a system from jamming, it should have a mapping to some other measure of distance to jamming, such as $\phi_m-\phi$. We shall explore this in the following section.
}
% With decreasing \mumicro, the average coordination number $Z$ increases for any given \Iv value. Hence, at the \Iv regime where \muofiv plateaus, we find increased $Z$ values. This also makes sense in terms of jamming volume fraction values (\phim), as \phim increases with diminishing \mumicro values, and higher $Z$ values should be expected at higher \phim values. The change in \phim essentially influences the distance to jamming ($\phi_m-\phi$), as does the changing shear stresses (\shear) and \mumicro in the constant $\phi$ system we use. We shall discuss \mumacro in terms of \shear, \mumicro and $\phi_m-\phi$ in \cref{section_distance_to_jamming}.

\subsection{Macroscopic friction coefficient and distance to jamming} \label{section_distance_to_jamming}

In the simulations, a range  of shear stresses (\shear), volume fractions ($\phi$) and microscopic friction coefficients (\mumicro) are studied. From previous experiments and simulations  \cite{singh2018constitutive,wyart2014discontinuous}, we understand the effect of changing each of these parameters on the rheology, especially on the jamming volume fraction (\phim). 
\hlo{Shear thickening is due to the formation of system spanning frictional networks, and the best way to describe this is to look at the fraction of frictional particles in the system.}
Beyond a characteristic shear stress $\sigma_0$, the fraction of particles in the system that have frictional contacts ($f$) increases until all particles become frictional \cite{Mari2014b}. This increase in $f$ with shear stress $\sigma_\mathrm{shear}$ can be described \cite{singh2017microstructural} as
\begin{align}
\sigma_0 &= F_\mathrm{CL}/6\pi R^2 \label{eq_sigma0} \\
\tilde\sigma &= \sigma_\mathrm{shear}/\sigma_0 \label{eq_sigmatilde} \\
f&=e^{(-1.45/\tilde\sigma)} \label{eq_f} 
\end{align}
\hlo{where R is the average radius of the particles, $F_\mathrm{CL}$ is the onset normal force between particles to initiate friction, and $\sigma_0 = F_\mathrm{CL}/(6 \pi R^2)$ is the characteristic stress for the onset of friction.}
Increasing the fraction of frictional particles leads to a lower jamming volume fraction \phim, as \phim for frictional particles is lower than non frictional particles \cite{wyart2014discontinuous,singh2018constitutive}. 
\hlo{This is a result of the frictional particles requiring a smaller number of inter-particle contacts to be arrested in comparison with frictionless particles~\cite{song2008phase}. The average coordination number for jamming ($Z_J$) in suspensions varies continuously between $Z_J(\mu_m=\infty)=4$ and $Z_J(\mu_m=0)=6$ in suspensions. 
% Changing the fraction of frictional particles in the system essentially has the effect of varying the jamming volume fraction $\phi_m$ between $\phi_J$(the jamming volume  fraction in a fully frictional system which is function of the microscopic friction coefficient), and $\phi^0_J$ (the jamming volume fraction in a frictionless system). 
Increasing the fraction of frictional particles in the system reduces the jamming volume fraction $\phi_m$ from that of a lubricated, non-frictional suspension ($\phi^0_J$) to that of a frictional suspension ($\phi_J$). $\phi_J(\mu_m)$ is the jamming volume fraction in a suspension with all particles in frictional contact and is a decreasing function of the microscopic friction coefficient \mumicro.  
% Increasing the microscopic friction coefficient \mumicro reduces $\phi_J(\mu_m)$. 
Hence, the volume fraction associated with jamming varies with \mumicro and the fraction of frictional particles $f$ in the system, and can be described \cite{singh2018constitutive} by}
\begin{align}
\phi_m(\tilde\sigma,\mu_m) &= \phi_J(\mu_m)f(\tilde\sigma) + \phi^0_J(1-f(\tilde\sigma)) \label{eq_phim}
\end{align}
where $\phi_J(\mu_m)$ represents the jamming volume fraction when $f=1$ for a given microscopic friction coefficient \mumicro. $\phi^0_J$ is the jamming volume fraction when $f=0$, which is equivalent to a $\mu_m=0$ (frictionless) state. 
Changing the microscopic friction coefficient  \mumicro influences \phim, as lowering \mumicro increases $\phi_J$, according to \cref{eq_phimu} \cite{singh2018constitutive}
% Hence, changing \shear, \mumicro and $\phi$ influences $\phi_m - \phi$, the distance to jamming. These effects on \phim can be modeled as follows:
\begin{align}
\phi_J(\mu_m) &= \phi_J^0 - ( \phi_J^0 - \phi^{\infty}_J)e^{-\mu_{\phi}/\mu_m} \label{eq_phimu} 
\end{align}
Here, $\phi^\infty_J$ is the jamming volume fraction at large \mumicro values, and $\mu_\phi$ is a constant. 
% It should be noted that $f$ and \phim are kept constant in the constant pressure experiments ($\phi_m$ can be reduced by increasing $P$, as evidenced by the increased \shear at similar \yd ranges shown in Fig:2(a) of \cite{boyer2011unifying}). 
Boyer et al. \cite{boyer2011unifying}  proposed a model for \Iv in terms of \phim and $\phi$ as:
\begin{align}
\phi(I_v)  = \dfrac{\phi_m}{1+I_v^{0.5}}
\label{eq_Iv_phi}
\end{align}
when substituted in \cref{eq_muofiv}, this gives \mumacro as a function of \phim and $\phi$:
\begin{align}
\mu(\phi,\phi_m) = \underbrace{\mu_1 + \dfrac{\mu_2-\mu_1}{1+I_0 \phi^2/(\phi_m-\phi)^2}}_{\mu_c} + \underbrace{ \left(\dfrac{\phi_m-\phi}{\phi}\right)^2 + \dfrac{5}{2}\dfrac{\phi_m}{\phi}(\phi_m-\phi)}_{\mu_h}
\label{eq_muofphi}
\end{align}

% Under constant pressure settings, regardless of the shear rate imposed,  the average normal force between particles is maintained constant. This implies that the fraction of particles that have contact normal forces that exceeds the critical load $F_{CL}$ and becomes frictional is also a constant. Assuming a constant \mumicro value, this results in a constant \phim value for a given imposed pressure $P$ (see \cref{eq_phim}). 
Under constant volume settings, the fraction of the frictional contacts varies with shear stress (or shear rate) in the system, which in turn varies \phim. 
% Hence, in constant pressure systems $\phi$ varies while \phim is constant, while in constant volume systems \phim varies while $\phi$ is constant. 
% Hence, in constant volume systems \phim varies while $\phi$ is constant. 
We can account for this variation in \phim by employing \cref{eq_sigma0,eq_sigmatilde,eq_f,eq_phim,eq_phimu}.
This helps to predict \phim in our constant volume system in terms of \shear and \mumicro which in turn enables an analysis of  \mumacro as a function of \phim-$\phi$ (i.e. a distance to jamming metric) and compare against the predictions from \cref{eq_muofphi}.

\begin{figure}[t]
\centering
\includegraphics[width=150mm]{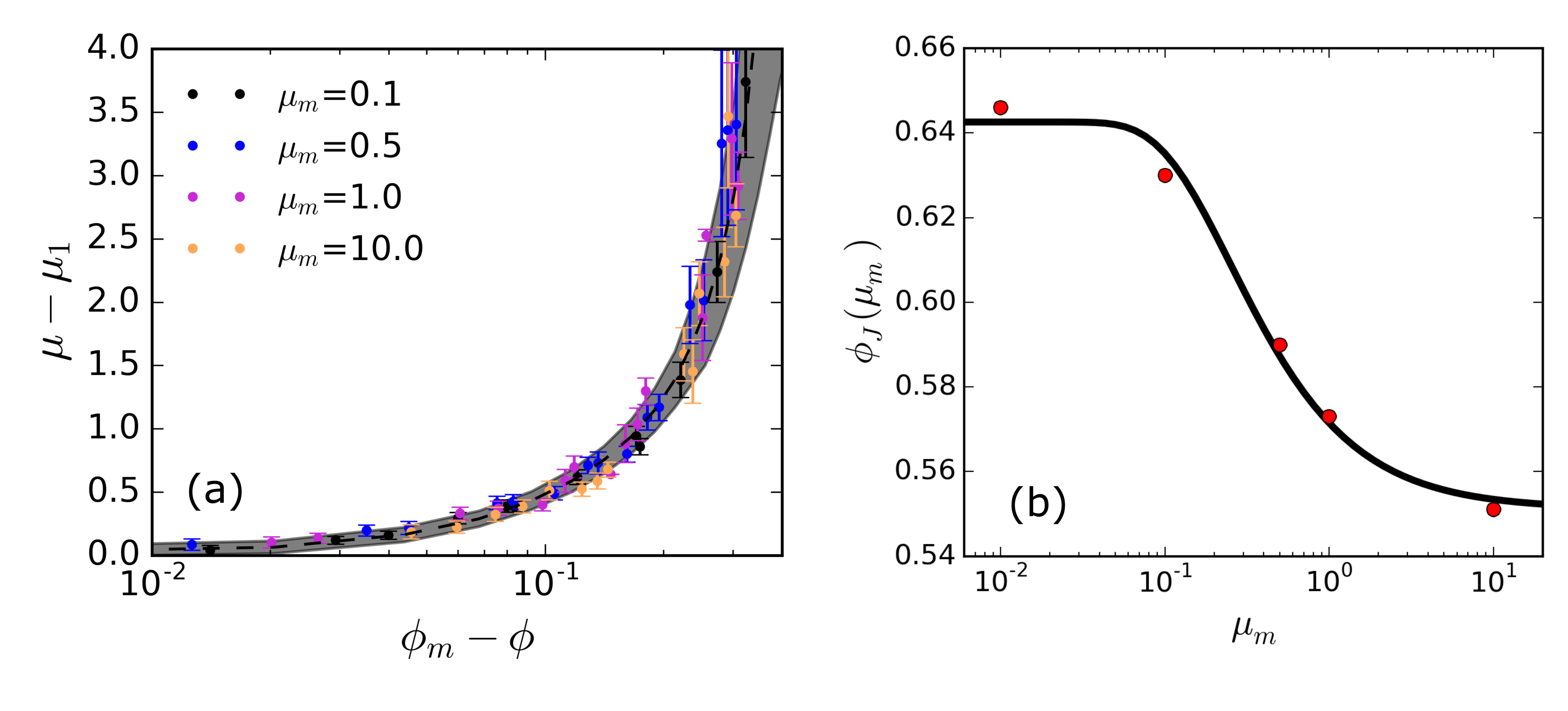}
% \decoRule
\caption{\textbf{(a)} Macroscopic friction coefficient $\mu(\phi,\sigma_\mathrm{shear},\mu_m)-\mu_1(\mu_m)$  vs distance to jamming $\phi_m - \phi$ for different \mumicro, \shear, and $\phi$ values. Shaded area represents the range of values of $\mu-\mu_1$ predicted by \cref{eq_muofphi}, correcting for changes in \phim according to \cref{eq_sigma0,eq_sigmatilde,eq_f,eq_phim,eq_phimu}, and the dashed line represents their mean. $\mu_1$ values are as given by \cref{fig_mu_vs_Iv_mus_all}(b), $\mu_2=0.7$. \textbf{(b)} Frictional jamming volume fraction $\phi_J(\mu_s)$ for different microscopic friction coefficient ($\mu_m$) values. Red dots represent the simulation data, while the curve represents the model presented in \cref{eq_phimu} with $\phi^0_J=0.643$, $\phi^\infty_J=0.55$ and $\mu^\phi = 0.25$.}
\label{fig_mu_vs_distance_to_jamming}
\end{figure} 

\cref{fig_mu_vs_distance_to_jamming}(a) shows the $\mu-\mu_1$ as a function of \phim - $\phi$ compiled over a range of \shear, $\phi$ and \mumacro values. The simulation results show agreement with the predictions from theory outlined in \cref{eq_muofphi,eq_sigma0,eq_sigmatilde,eq_f,eq_phim,eq_Iv_phi,eq_phimu}. The changes in $\phi_J$ with \mumicro are taken into account by using their relationship outlined in \cref{eq_phimu}, as shown in \cref{fig_mu_vs_distance_to_jamming}(b). The simulation results agree with the theoretical assumption  that, by accounting for changes in $\phi_m$ with \shear and \mumicro, the values of \mumacro across different \shear and \mumicro values  collapse to the regime outlined in \cref{fig_mu_vs_distance_to_jamming}(a). The change in the frictional jamming volume fraction $\phi_J$ with \mumicro is shown in \cref{fig_mu_vs_distance_to_jamming}(b), along with the model presented in \cref{eq_phimu}. 
\hlo{The results also show that $\mu-\mu_1$ is indeed a measure for the distance to jamming, as suggested in the previous section.}

\subsection{Microstructure changes} \label{section_microstructure}
The microscopic friction coefficient plays an important role in the nature of contact networks formed at jamming. The mean coordination number at which the suspension jams ($Z_J$), is inversely dependent on \mumicro, as  $Z_J(\mu_m=0)=6$ and $Z_J(\mu_m=\infty)=4$ \cite{song2008phase}. The evolution of  \mumacro with average coordination number ($Z$) under varying \mumicro values thus, is of interest. It is also compelling to view \muofiv rheology in terms of the evolution of $Z$.

\begin{figure}[h]
\centering
\includegraphics[width=150mm]{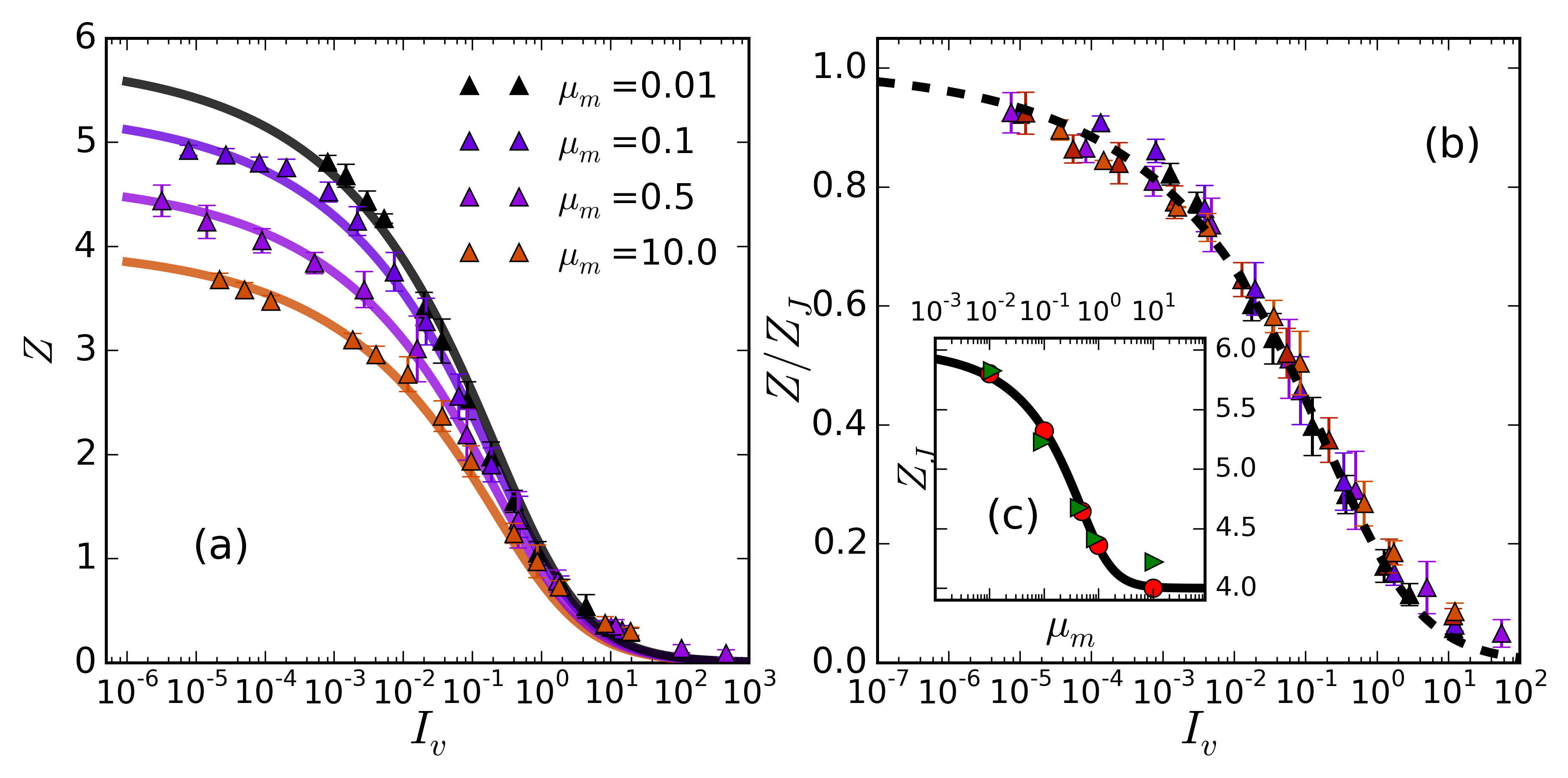}
% \decoRule
\caption{\textbf{(a)} Average coordination number $Z$  as a function of  viscous number $I_v$ for different \mumicro compiled across different $\phi$ and \shear values. Each dot corresponds to simulation results at corresponding \mumicro. Lines show the prediction of $Z(I_v)$ from \cref{eq_ZbyZJ_Iv,eq_ZJ_mum}. \textbf{(b)} $Z$ normalized by jamming coordination number $Z_J$ vs $I_v$. The dashed line represents the $Z/Z_J(I_v)$ model from \cref{eq_ZbyZJ_Iv} while dots represent the simulation results of \mumicro. \textbf{(c)} Variation in $Z_j$ with \mumicro. The dots show $Z_J$ as observed in simulations at vanishing $I_v$. The line represents the $Z_J(\mu_m)$ model from \cref{eq_ZJ_mum}. Green triangles represent the random loose packing limits in simulations of granular system \cite{song2008phase}.}
\label{fig_Z_vs_Iv}
\end{figure} 

\cref{fig_Z_vs_Iv}(a) shows average coordination number  $Z(I_v)$ under various \mumicro values. $Z$ is calculated per particle by counting the number contacts it makes, i.e. cases where $r_{ij}-R_i-R_j \leq d_c$ where $r_{i,j}$ are the distance between the particles and and $R_{i,j}$ are their radii. Even though the data is compiled from various $\phi$ and \shear values, $Z(I_v,\mu_m)$ collapses to  unique curves depending on \mumicro. The maximum coordination number is $Z \approx 4$ at $\mu_m=10.0$ and saturates at higher maximum values ($Z_J$) with reducing \mumicro as expected from $Z_J(\mu_m)$ relationship described before. The low $Z$ values at large \Iv sheds light on the the insensitivity of \muofiv rheology to changes in \mumicro in these \Iv ranges. \muofiv rheology hence is essentially the process of varying  coordination numbers between zero and $Z_J(\mu_m)$.   
Upon normalizing $Z$ by $Z_J(\mu_m)$, the different $Z(\mu_m)$ curves collapse to a single curve, which can be modeled as:
\begin{align}
\frac{Z}{Z_J} = 1 - (1+I_v^{\alpha_1})^{-\beta_1}
% \frac{Z}{Z_J} = \dfrac{1}{1+\alpha_1 I_v^{\beta_1}}
\label{eq_ZbyZJ_Iv}
\end{align}

where $\alpha_1 = 0.77$ and $\beta_1=0.3$.
% where $\alpha_1=3.3$ and $\beta_1=0.4$.
The variation in $Z_J$ between 6 and 4 depending on \mumicro can also be modeled using the expression:
\begin{align}
Z_J = 6 - 2(1+\mu_m^{\alpha_2})^{-\beta_2}\label{eq_ZJ_mum}
\end{align}
where $\alpha_2 = -1.72$ and $\beta_2=0.27$. \cref{fig_Z_vs_Iv}(b) shows $Z/Z_J$ as a function of \Iv, and it can be observed that the data collapses to a single curve, modeled by \cref{eq_ZbyZJ_Iv}. The variation in $Z_J$ with \mumicro, modeled by \cref{eq_ZJ_mum} is shown in \cref{fig_Z_vs_Iv}(c).  
\hlo{It is relevant to note that the variation in $Z_J$ with \mumicro is found to be quite similar to the change in the coordination numbers associated with minimum random loose packing (RLP) limit observed in dry granular systems~\cite{song2008phase}. The minimum RLP coordination number corresponds to the minimum coordination number required to obtain a disordered, mechanically stable jammed system. As the limits of jamming are prescribed entirely by the properties of the particles, it is conceivable that the characteristics related to jamming in granular systems devoid of fluid is to be expected in suspensions as well.}

\begin{figure}[h]
\centering
\includegraphics[width=150mm]{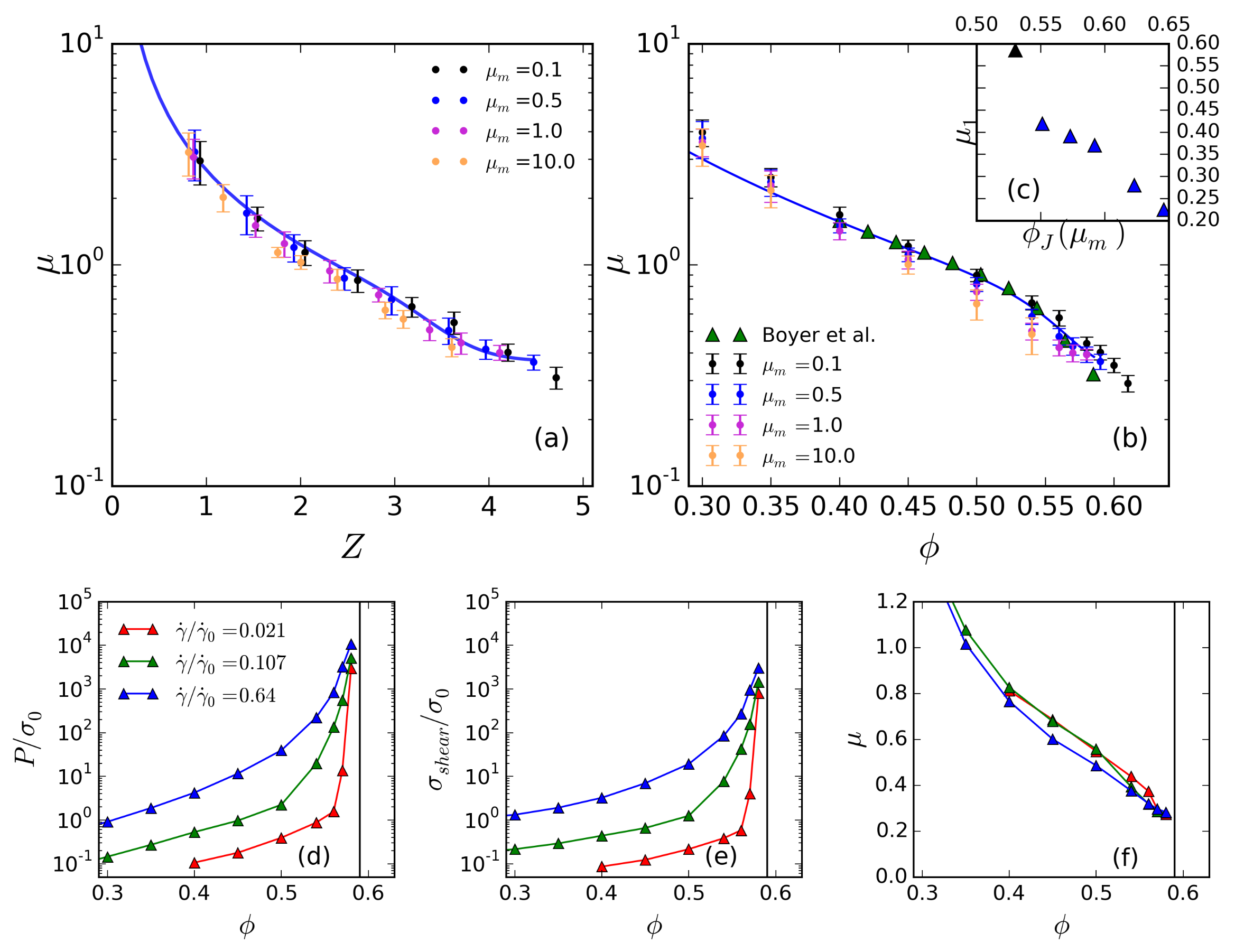}
% \includegraphics[width=150mm]{paper3_Z_vs_mumacro.png}
% \includegraphics[width=150mm]{pressure_shear_scaling_with_phi.png}
% \decoRule
\caption{\textbf{(a)} Variation in the macroscopic friction coefficient \mumacro with average coordination number $Z$  compiled across different \mumicro and $\phi$ values. Solid line represents the theoretical prediction of $\mu(Z)$ for $\mu_m=~0.5$.
\textbf{(b)} Variation of \mumacro with $\phi$ for various \mumicro values. Solid line represents the theoretical prediction of $\mu(\phi)$ for $\mu_m=~0.5$. Green triangles represent the experimental results from Boyer et al. \cite{boyer2011unifying}. 
\textbf{(c)} Minimum macroscopic friction coefficient ($\mu_1$) achieved at jamming as a function of the jamming volume fraction $\phi_J(\mu_m)$. Black triangle represents the $\mu_1$ observed at jamming for suspensions of non-spherical particles discussed in \cref{section_nonsperhical_particles}.
\textbf{(d,e)} Pressure $P$ and shear stress $\sigma_\mathrm{shear}$ normalized by $\sigma_0$ scaling with volume fraction for various shear rates for $\mu_m=0.5$.  
\textbf{(f)} Macroscopic friction coefficient \mumacro measured for the pressure and shear stresses shown in \textbf{(d,e)}. Vertical black lines show the jamming volume fraction.
}
\label{fig_Z_vs_mumacro}
\end{figure}

The effect of changing $Z$ on \mumacro, under various \mumicro values is shown in \cref{fig_Z_vs_mumacro}(a).  $\mu(Z)$ values reasonably collapses into a single curve for all values of \mumicro studied.  This demonstrates that the the minimum \mumacro achieved at low \Iv values (i.e. $\mu_1$) is determined by $Z_J$. As $Z_J$ is inversely related to \mumicro, the relationship between $\mu_1$ and \mumicro depicted in \cref{fig_mu_vs_Iv_mus_all}(b) can be rationalized. Assuming a range of $I_v$ values, one can calculate and compare \mumacro against $Z$ for a given \mumicro value using the relationships outlined in \cref{eq_ZJ_mum,eq_ZbyZJ_Iv,eq_phimu,eq_muofiv}. As shown in \cref{fig_Z_vs_mumacro}(a), the theoretical predictions of $\mu(Z,\mu_m=0.5)$ is in agreement with the simulation results.  
\hlo{Consequently, the variation in \mumacro with $\phi$ also collapses reasonably onto a simple curve across the various \mumicro values studied, as seen in \cref{fig_Z_vs_mumacro}(b). This behavior is observed in 2D simulations of sheared suspensions and dense granular systems \cite{thomas2018microscopic,da2005rheophysics} and experimentally by Boyer et al.\cite{boyer2011unifying}. 
With increasing volume fraction, under a given shear rate, the  shear stress and normal stresses become larger, but their ratio (\mumacro) reduces till $\mu=\mu_1$ at jamming (see \cref{fig_Z_vs_mumacro}(d-f)). 

This implies that the jamming volume fraction determines $\mu_1$, the minimum macroscopic friction coefficient. The lower the jamming volume fraction, the higher the observed $\mu_1$; see \cref{fig_Z_vs_mumacro}(c). Our simulations of non-spherical particle suspensions (see next section) that jam at a lower volume fraction compared to spherical particles also agree with this observation, as shown in \cref{fig_Z_vs_mumacro}(c).
}

\subsection{Non spherical particles}
\label{section_nonsperhical_particles}
\begin{figure}[h]
\centering
\includegraphics[width=140mm]{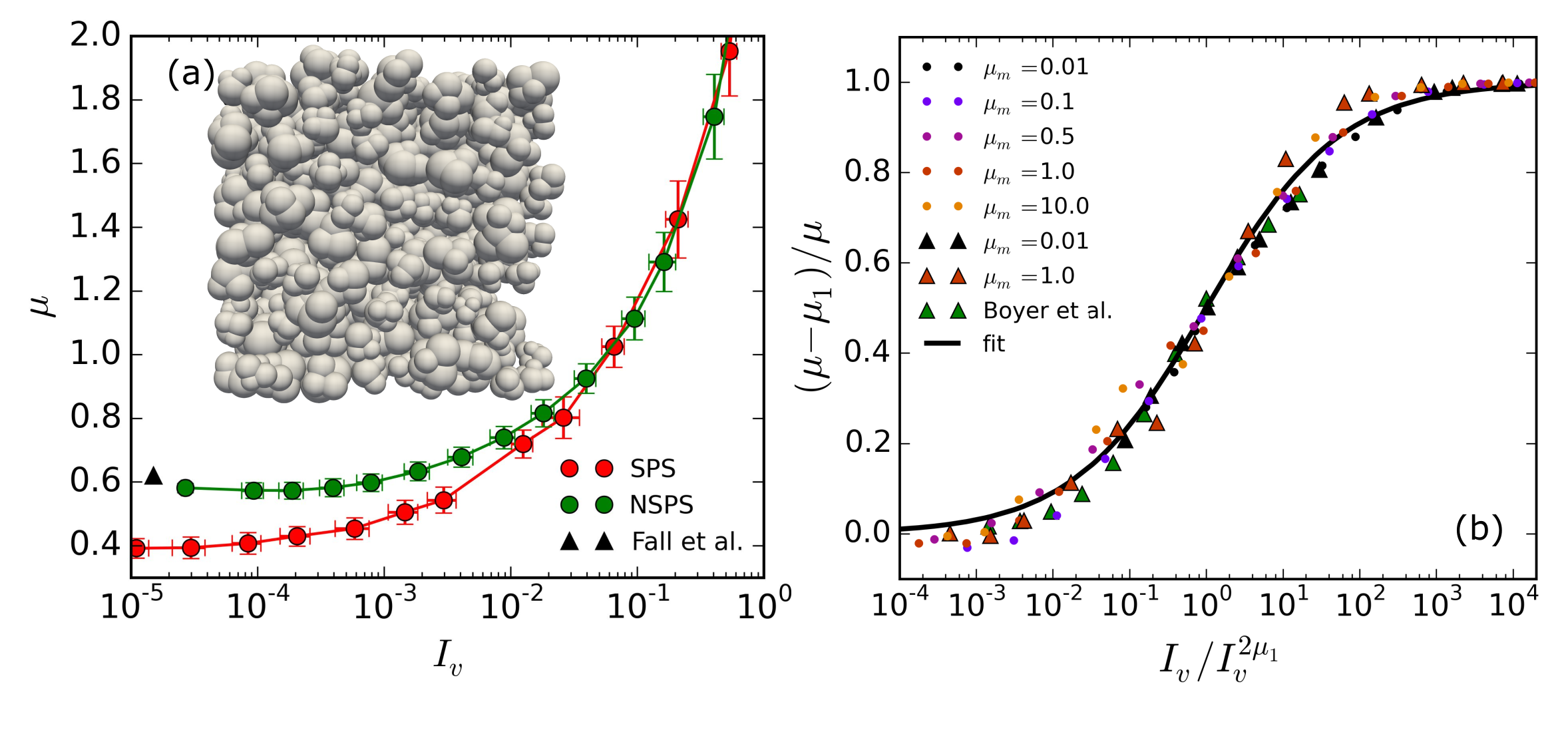}
% \decoRule
\caption{\textbf{(a)} Macroscopic friction coefficient $\mu$  vs  viscous number \Iv for spherical particle suspensions (SPS) and non spherical particle suspensions (NSPS) with \mumicro=1.0. Black triangle represent the macroscopic friction coefficient measured close to jamming in experiments with cornstarch suspensions \cite{Fall2012a}. \textbf{(inset)}:Representation of the non-spherical particles used in the simulations. \textbf{(b)}$(\mu-\mu_1)/\mu$ for spherical (dots) and non-spherical (triangles) particle suspensions. Green triangles represent the results of Boyer et al.\cite{boyer2011unifying}. Line represents the fit given by \cref{eq_mu_minus_mu1_by_mu1_Iv_I2v_all}.}
\label{fig_mu_vs_Iv_spherical_nonspherical}
\end{figure}

Particle shapes have significant effects on the shear thickening behavior of the suspensions. Cornstarch particles are observed to shear thicken at much lower \phim values (\phim $\approx 0.44$) \cite{Fall2012a} in comparison to suspensions of spherical particles which shear thicken around $\phi_m=0.56$. Simulation results \cite{Lorenz2018} show that frictional jamming volume fraction $\phi_J^\infty$ is lowered when particles shapes become 'cornstarch-like'. In the interest of comparing the macroscopic friction coefficient variation in spherical particles to that of non-spherical particles, simulations of  'cornstarch-like' non-spherical particle suspensions were performed.  The 'cornstarch-like' particles were created using overlapping spheres of varying sizes, as outlined in \cite{Lorenz2018}. A representation of the non-spherical particles used is provided in \cref{fig_mu_vs_Iv_spherical_nonspherical}(a)(inset). 

\cref{fig_mu_vs_Iv_spherical_nonspherical}(a) compares \muofiv for spherical particle suspensions and non-spherical particle suspensions. At high viscous numbers, \muofiv for spherical and non-spherical particle suspensions tends to be the same. This is understandable, as at high \Iv values the coordination numbers of the particles (spherical or non-spherical) in the suspensions reduces and particle shapes become increasingly less relevant. However, at small \Iv values, \muofiv behavior of non-spherical particle suspensions deviates from that of spherical particle suspensions, for any constant \mumicro value. Naturally, these deviations become apparent at \Iv values where particle interactions become relevant, i.e. $I_v < 10^{-1}$. Results suggests that the macroscopic friction coefficient of non-spherical particle suspensions plateaus to $\mu_1$ at higher viscous numbers in comparison to the spherical particle suspensions. Also, at vanishing viscous numbers, the macroscopic friction coefficient of the non-spherical particle suspensions saturates to a higher $\mu_1$ in comparison with spherical particle suspensions, for a given \mumicro value.
\hlo{This agrees with measurements of the macroscopic friction coefficient for cornstarch suspensions  close to jamming \cite{Fall2012a}, where $\mu_1\approx0.62$ in the experimental systems and $\mu_1\approx0.6$ in the simulations. In the previous section, it was concluded that the jamming volume fraction determines the minimum value of the macroscopic friction coefficient. Considering that the non-spherical suspension simulated here jams around $\phi^{non-spherical}_J=0.53$, which is lower than the jamming volume fraction for spherical particles ($\phi^{spherical}_J=0.576$) at the same \mumicro value ($\mu_m=1$), the larger $\mu_1$ observed here can be rationalized.

It is intriguing to see whether one can generalize these variations in \mumacro with particle shapes and microscopic friction coefficients to arrive at a common curve for all available data. By (a) normalizing \Iv with $I_v^{2\mu_1}$ (where $I_v^{2\mu_1} = I_v(\mu=2\mu_1)$) to account for the shift in \Iv values at which \mumacro plateaus to $\mu_1$, and  (b)  setting upper and lower bounds to the variation in \mumacro by using $(\mu-\mu_1)/\mu$ as the measure of the variation of \mumacro with \Iv, the results collapses nicely to a single curve, for both spherical and non-spherical particle suspensions, across varying $\mu_m$ values (see  \cref{fig_mu_vs_Iv_spherical_nonspherical}(b)). The results of Boyer et al. \cite{boyer2011unifying} are shown for comparison, and also agrees with the curve. This common relationship can be fitted using the curve given by:
\begin{align}
    \frac{\mu-\mu1}{\mu} = \dfrac{\sqrt{I_v}}{\sqrt{I_v}+\sqrt{I^{2\mu_1}_v}}
    \label{eq_mu_minus_mu1_by_mu1_Iv_I2v_all}
\end{align}
which in turn gives:
\begin{align}
    \mu = \mu_1 \left (1+\sqrt{\frac{I_v}{I^{2\mu_1}_v}}\right )
    \label{eq_mu_mu1_Iv_I2v_all}
\end{align}

Even though the simulation results conform to the expression given by \cref{eq_mu_minus_mu1_by_mu1_Iv_I2v_all}, it should be mentioned that  the validity of the expression at high viscous numbers ($I_v>0.5$) is suspect, as we have no experimental data in this regime. Experimental data for non-spherical particles at viscous numbers high enough to obtain $I^{2\mu_1}_v$ is also absent, which prevents us from further validation.      
}

% \begin{figure}[h]
% \centering
% \includegraphics[width=165mm]{paper3_spherical_vs_nonsperical_mu_vs_Z_Iv.png}
% % \decoRule
% \caption{\textbf{(a)} Average coordination number $Z$  vs viscous number \Iv  \textbf{(b)} Macroscopic friction coefficient \mumacro vs $Z$ for spherical (SPS) and non spherical particle (NSPS) suspensions with \mumicro=1.0.}
% \label{fig_mu_vs_Iv_Z_spherical_nonspherical}
% \end{figure} 

% \cref{fig_mu_vs_Iv_Z_spherical_nonspherical}(a) shows the scaling of the average coordination number $Z$ with \Iv for SPS and NSPS. Coordination number of an individual non-spherical particle is calculated as the sum of the contacts made by each individual sub-spheres that constitutes the non-spherical particle. The 'cornstarch-like' NSPS used here used has higher coordination numbers  at same \Iv values, while also differing in the scaling of $Z(I_v)$ in comparison to SPS. It is possible that the scaling of $\phi_m$ with \mumicro is  different for NSPS compared to SPS.  The $\mu(Z)$ behavior shown in \cref{fig_mu_vs_Iv_Z_spherical_nonspherical}(b) shows higher coordination numbers at similar \mumacro values.  
% \textbf{(Comparison with Experiments, to be added)}
% \subsection{Comparison with Experiments} \label{section_comparison_with_experiments}

% \input{Conclusion_and_Discussion}
\section{Conclusion}
\hlo{
We analyze the behavior of the macroscopic friction coefficient (\mumacro) under different microscopic friction coefficients (\mumicro) using 3D numerical simulations. The predictions of \mumacro from simulations agree with earlier predictions of viscous number granular suspension rheology. 
 We find that when $\mu_m>0.3$, that viscous number rheology is largely insensitive to the value of \mumicro. 
By changing the jamming volume fraction $\phi_m$ with the changes in shear stresses and \mumicro,  we analyze \mumacro in terms of distance to jamming ($\phi_m - \phi$) and provide phenomenological but analytic formulae that match the observations.
Our results also suggest the behavior of \mumacro across various \mumicro and viscous numbers (\Iv) can be reduced to effects of distance to jamming. 
The study of changes in the average coordination number ($Z$) with viscous number (\Iv) shows that $Z$ smoothly decreases from $Z_J$ ($Z$ at jamming) to zero with increasing viscous number, where $Z_J$ is again determined by \mumicro.
Our results suggest that the minimum \mumacro achieved is inversely related to the jamming volume fraction and $Z_J$.
Finally, we show that with appropriate scaling, a common curve for the variation of \mumacro with \Iv emerges for both spherical and non-spherical particles under varying \mumicro values.    
% Our observations suggest that the minimum \mumacro achieved close to jamming is determined by the maximum  particle volume fraction attained for a given \mumicro value. This provides a rationale for the reducing minimum \mumacro close to jamming with reducing \mumicro values. Finally, we show that by appropriately scaling \mumacro and \Iv, a common curve for their variation emerges for both spherical and non-spherical particles under varying \mumicro values.    
}
% We have analyzed the behavior of the macroscopic friction coefficient in 3D simulations of shear thickening suspension. We  we observe that the macroscopic friction coefficient decreases as $\phi$ increases. We show that the simulation results conforms to $\mu(I_v)$ rheology, despite the simulations being constant $\phi$ systems. We also perform simulations of varying microscopic (inter-particle) friction coefficient $\mu_m$ and find the the macroscopic friction coefficient \textit{increases} if $\mu_m$ is decreased. In order to rationalize this observations and incorporate the effect of the change of $\mu_m$ on behavior of the $\mu$, we incorporate the change in $\phi_m$ with $\mu_m$ into the $\mu(I_v)$ rheology, and find good agreement with the simulations. 

\section*{Conflicts of Interest}
There are no conflicts of interests.
	
\section*{ACKNOWLEDGEMENTS}
 We would like to thank SURFsara for using their HPC infrastructure and for providing support (project number 00231267). Author VS acknowledges funding by NWO, Netherlands under the CSER program (project number 14CSER026). 
%\section*{Bibliography}
\bibliography{bibliography}

\end{document}